\documentstyle[preprint,aps,version2]{revtex}
\begin{document}
\draft
\rightline{BIHEP-TH-94-45}
\begin{title}

Re-examination of the Perturbative Pion Form Factor with \\
	Sudakov Suppression
\end{title}

\author {Fu-guang Cao, Tao Huang and Chuan-wang Luo}

\begin{instit}
	 Institute of High Energy Physics, Academia Sinica,
	 P.O.Box 918, Beijing, 100039,
           P. R. China\footnote{E-mail address:Huangt@bepc2.ihep.ac.cn}
\end{instit}

\begin{abstract}

The perturbative pion form factor with Sudakov suppression is re-examined.
Taking into account the multi-gluon exchange in the law $Q^2$ regions,
we suggest that the running coupling constant should be frozen at
$\alpha_s(t=\sqrt{<{\bf k}_{T}^{2}>})$ and $\sqrt{<{\bf k}_{T}^{2}>}$
is the average transverse momentum which can be determined by the
pionic wave function.
In addition,
we correct the previous calculations about the Sudakov
suppression factor which plays an important role in
the perturbative predictions for the pion
form factor.
\end{abstract}

\pacs{ PACS numbers: }

\pagestyle{plain}

It is believed that perturbative QCD (pQCD) can successfully
describe exclusive
processes at asymptotically large momentum transfers \cite{BEM}.
However, the
applicability of the pQCD to the pion form factor at the present
energy is a matter of controversy \cite{Isgur}.
Recent studies [3-5]
on the pion electronmagnetic form factor show that the
pQCD contributions become self-consistent for momentum
transfers at a few GeV range. Li and Sterman \cite{Hsiang} give
a modified expression
for the pion form factor by taking into account the
customarily neglected partonic transverse momenta as well as the
Sudakov corrections. Jakob and Kroll \cite{Jakob} point out that the
dependence of the hadronic wave function on the intrinsic
transverse momentum
should be considered in the perturbative calculation.
Sudakov corrections come from an infinite
summation of higher-order effects associated with the elastic
scattering of the valence partons.
However, because the running coupling constant $\alpha_{s}$
becomes rather large with $b$
(the distance between a quark-antiquark pair)
increasing in the end-point regions, a cut-off on
$\alpha _{s}$ has to be made to evaluate perturbative contributions
and to justify the self-consistency of perturbative calculations.
In this paper, we re-examine the perturbative pion form factor with
the Sudakov suppression. It is pointed out that $\alpha _{s}(t)$ should
be frozen as $t$ is smaller than a certain value due to the
multi-gluon exchange at low $Q^2$. We suggest that the frozen point
is related to the average  transverse momentum
which is determined by the pionic wave function.
In addition,
we correct the previous calculations about the Sudakov
suppression factor which plays an important role in
the perturbative predictions for the pion
form factor.

Let us begin with a brief review on the derivation of the
expression for the pion form factor in Ref. \cite{Hsiang}
Taking into account
the transverse momenta ${\bf k}_{T}$ that
flow from the wave functions through the hard scattering
leads to a factorization form with two wave
functions $\psi_{i}(x_{i},{\bf k}_{T_{i}})$ corresponding
to the external pions, combined with a new hard-scattering
function $T_{H}(x_{1},x_{2},Q,{\bf k}_{T_{1}},{\bf k}_{T_{2}})$,
which depends in general on transverse as well as longitudinal
momenta,
$$F_{\pi}(Q^{2})=\int_0^1 dx_{1}dx_{2}\int d^{2}{\bf k}_{T_{1}}
d^{2}{\bf k}_{T_{2}}\psi_{i}(x_{1},{\bf k}_{T_{1}},P_{1})$$
$$~~~~~~~~\times T_{H}(x_{1},x_{2},Q,{\bf k}_{T_{i}},\mu)
\psi_{i}(x_{2},{\bf k}_{T_{2}},P_{2}),\eqno(1)$$
where $Q^{2}=2P_{1}\cdot P_{2}$, and $\mu$ is the renormalization
and factorization scale.

Through Fourier transformation eq. (1) can be expressed as
$$F_{\pi}(Q^{2})=\int_0^1 dx_{1}dx_{2}\frac{d{\bf b}_{1}}{(2\pi)^{2}}
\frac{d{\bf b}_{2}}{(2\pi)^{2}}
\varphi(x_{1},{\bf b}_{1},P_{1},\mu)$$
$$~~~~~~~~~~~~\times T_{H}(x_{1},x_{2},Q,{\bf b}_{1},{\bf b}_{2},\mu)
\varphi(x_{2},{\bf b}_{2},P_{2},\mu).\eqno(2)$$
In this expression, wave function $\varphi(x_{i},{\bf b}_{i},P_{i},\mu)$
takes into account
an infinite
summation of higher-order effects associated with the elastic
scattering of the valence partons, which gives out the Sudakov
suppression to the large-$b$ and small-$x$ regions. At the same
time the
intrinsic transverse momentum dependence of the wave function
provides another suppression to the large-$b$ regions.

At the lowest order, $T_{H}$ is given by
$$T_{H}(x_{1},x_{2},Q,{\bf k}_{T_{i}})=
\frac{16\pi C_{F} \alpha_{s}(\mu)}{x_{1}x_{2}Q^{2}
+({\bf k}_{T_{1}}+{\bf k}_{T_{2}})^{2}},\eqno(3)$$
where $C_{F}$ is the color factor. The wave function
can be expressed as
$$\varphi(x,b,P,\mu)=\exp\left[ -s(x,b,Q)-s(1-x,b,Q)
-2 \int_{1/b}^{\mu}\frac{d\bar{\mu}}{\bar{\mu}}\gamma_{q}
(g(\bar{\mu}))\right]\times \phi\left(x,\frac{1}{b}\right),\eqno(4)$$
where $\gamma_{q}=-\alpha_{s}/\pi$ is the quark anomalous dimension in
the axial gauge.
$s(\xi,b,Q)$ is the Sudakov exponent factor, which reads
$$s(\xi,b,Q)=\frac{A^{(1)}}{2\beta_{1}}\hat{q}~
\ln\left(\frac{\hat{q}}{-\hat{b}}\right)+\frac{A^{(2)}}{4\beta_{1}^{2}}
\left(\frac{\hat{q}}{-\hat{b}}-1\right)-\frac{A^{(1)}}{2\beta_{1}}
(\hat{q}+\hat{b})$$
$$~~~~~~~~~~ -\frac{A^{(1)}\beta_{2}}{4\beta_{1}^{3}}\hat{q}
\left[\frac{\ln(-2\hat{b})+1}{-\hat{b}}-\frac{\ln(-2\hat{q})+1}
{-\hat{q}}\right]$$
$$~~~~~~ -\left(\frac{A^{(2)}}{4\beta_{1}^{2}}-
\frac{A^{(1)}}{4\beta_{1}}\ln(\frac{1}{2}e^{2\gamma-1})\right)
\ln\left(\frac{\hat{q}}{-\hat{b}}\right)$$
$$~~~~ +\frac{A^{(1)}\beta_{2}}{8\beta_{1}^{3}}
\left[\ln^{2}(2\hat{q})-\ln^{2}(-2\hat{b})\right],\eqno(5)$$
where
$$\hat{q}=\ln[\xi Q/(\sqrt{2}\Lambda)], ~~~~ \hat{b}=\ln(b\lambda),$$
$$\beta_{1}=\frac{33-2n_{f}}{12}, ~~~~~  \beta_{2}=\frac{153-19
n_{f}}{24},$$
$$A^{(1)}=\frac{4}{3},~~~~~~ A^{(2)}=\frac{67}{9}-\frac{1}{3}\pi^{2}
-\frac{10}{27}n_{f}+\frac{8}{3}\beta_{1}\ln(\frac{1}{2}e^{\gamma}),
\eqno(6)$$
where $n_{f}$ is the number of quark flavors and $\gamma$
is the Euler constant.

It should be noted that  there are  some mistakes
in the coefficients
of the fourth and the sixth terms in $s(\xi,b,Q)$
given by Refs. [4,6]. We find  that the correct coefficients
should  be $ -\frac{A^{(1)}\beta_{2}}{4\beta_{1}^{3}}$ and
$ +\frac{A^{(1)}\beta_{2}}{8\beta_{1}^{3}}$,
in place of $ -\frac{A^{(1)}\beta_{2}}{16\beta_{1}^{3}}$ and
$ -\frac{A^{(1)}\beta_{2}}{32\beta_{1}^{3}}$  in Refs. [4,6].
It is $s(\xi,b,Q)$  that play an important role
in the evaluation of the pion form factor.  In this paper,
we examine the effects  brought about by these corrections.

Applying  the renormalization group equation
to $T_{H}$ and substituting the
explicit expression for $T_{H}$, we  have the following expression
for the pion form factor
$$F_{\pi}(Q^{2})=16\pi C_{F}\int_{0}^{1}d x_{1}d x_{2}
\int_{0}^{\infty}b~db \alpha_{s}(t)K_{0}(\sqrt{x_{1}
x_{2}}Qb)$$
$$~~~~~~~\times \phi(x_{1},1/b)
\phi(x_{2},1/b) \exp (-S(x_{1},x_{2},Q,b,t),\eqno(7)$$
where
$$S(x_{1},x_{2},Q,b,t)=
\left[\sum_{i=1}^{2}\left(s(x_{i},b,Q)+
s(1-x_{i},b,Q)\right)-\frac{2}{\beta_{1}}\ln\frac{\hat{t}}
{-\hat{b}}\right].\eqno(8)$$

Radiative corrections in higher orders
will bring logarithms of the form $\ln(t/\mu)$ into $T_{H}$,
where $t$ is the largest mass scale appearing in $T_{H}$.
Ref. \cite{Hsiang} points out that a natural choice
for $\mu$ in $T_{H}$ is $\mu=t$ and
$$t=\max\left(\sqrt{x_{1} x_{2}} Q,1/b\right).\eqno(9)$$
If $b$ is
small, radiative corrections will be small regardless of the values
of $x$ because of the small $\alpha_{s}$. When $b$ is large
and $x_{1}x_{2}Q^{2}$ is small, radiative corrections are
still large in $T_{H}$, while $\varphi$ will suppress these regions.
But with $b$ increasing, $\alpha_{s}$ becomes rather large
(for example, $\alpha_s$ is large than unity when
$b$ is large than 0.5/$\Lambda_{QCD}$ GeV$^{-1}$ for $x_1=0.01$,
$x_2=0.01$ and $Q=2$ GeV, see Fig. 1)
and  accordingly the perturbative calculation loses its self-consistency.
Therefore, a cut-off on
$\alpha _{s}$ is made to evaluate perturbative contributions
and to justify the self-consistency of perturbative calculation.
That is to say, if $50\%$ of the result come from the regions where
$\alpha_{s}$ is not very large (say, $<0.7$),  the perturbative
calculation can be trusted.

Strictly speaking, the perturbative predictions to the regions
where $\alpha _{s}$ is larger than unity are unreliable, although
these regions are suppressed.
In fact, in the regions of small $x_1 x_2 Q^2$ and large $b$,
the multi-gluon exchange is important and the transverse momentum
intrinsic to the bound
state wave-functions flows through all the propagators.
To respect this point, instead of eq. (9)  we suggest that
$$t=\max\left(\sqrt{x_{1} x_{2}} Q,1/b_F \right),\eqno(10)$$
where
$$b_F=\left \{
	\begin{array}{cl}
	b~~~~~~~~~	& \mbox{if $1/b \geq \sqrt{<{\bf k}_{T}^{2}>}$ }\\
	1/\sqrt{<{\bf k}_{T}^{2}>}~~~~~~~~~ &
	\mbox{if $1/b< \sqrt{<{\bf k}_{T}^{2}>},$}
	\end{array}
	\right.
	\eqno(11)$$
where $\sqrt{<{\bf k}_{T}^{2}>}$ is the average transverse momentum.
With such a choice, the running coupling constant  will be
frozen at $\alpha_s(t=\sqrt{<{\bf k}_{T}^{2}>})$ when $b$ is large and
$x_1x_2Q^2$ is small.
In this way, the perturbative contributions
to the pion form factor  can be calculated from the present energy
with a reasonable $\alpha_{s}$,
and it should be emphasized that the average transverse momentum
$\sqrt{<{\bf k}_{T}^{2}>}$
always associates  with the hadronic wave function

\underline{{\it The pion wave function}}:
According to Brodsky-Huang-Lepage prescription \cite{BHL}, one can
connect the equal-time
wave function in the rest frame and the light-cone wave function by
equating the off-shell propagator in the two frames.
They got the wave function [7,3] at the infinite momentum frame
from the harmonic oscillator model at the rest frame \cite{Acta}
$$\psi^{(a)}(x,{\bf k}_T)=A~\exp\left[-\frac{{\bf k}_{T}^{2}+m^{2}}
{8 \beta^{2} x(1-x)}\right],\eqno(12)$$
where $m$=0.289 GeV, $\beta=0.385$ GeV, $A=32$ GeV$^{-1}$
are parameters which
are adjusted \cite{MA} by using the constraints derived \cite{BHL}
from $\pi \rightarrow \mu \nu$ and $\pi^0 \rightarrow \gamma \gamma$
decay amplitudes:
$$\int_0^1 dx \int \frac{d^2{\bf k}_T}{16\pi ^3}\psi(x,{\bf k}_T)=
  \frac{f_\pi}{2\sqrt{6}},\eqno(13)$$
$$\int_0^1dx \psi(x,{\bf k}_T=0)=\frac{\sqrt 6}{f_\pi},\eqno(14)$$
where $f_\pi=0.133$ GeV is
the pion decay constant.

The mean squared transverse momentum is defined as
$$<{\bf k}_{T}^{2}>=\int \frac{d^{2}{\bf k}_{T}}{16\pi^{3}}dx
|{\bf k}_{T}|^{2}|\psi(x,{\bf k}_{T})|^{2}/P_{q\bar{q}},\eqno(15)$$
where
$$P_{q\bar{q}}=\int \frac{d^{2}{\bf k}_{T}}{16\pi^{3}}dx
|\psi(x,{\bf k}_{T})|^{2}\eqno(16)$$
is the probability of finding $q\bar{q}$ Fock state in the pion.
For $\psi^{(a)}(x,{\bf k}_{T}), <{\bf k}_{T}^{2}>$=(0.356 GeV)$^{2}$.
Expressing  $\psi^{(a)}(x,{\bf k}_{T})$ in the $b$-space, we obtain
$$\phi^{(a)}(x,1/b)=\frac{2 A \beta^2}{(2 \pi)^2}
x(1-x)\exp\left(-\frac{m^{2}}{8\beta^{2}x(1-x)}\right)
\exp\left(-2\beta^2 x(1-x)b^2\right).\eqno(12')$$

In order to fit the experimental data and to suppress the end-point
contributions for the applicability of pQCD, a model for the pion
wave function has been proposed in Refs. [10,11] by simply
adding a factorizing function
$(1-2x)^{2}$ to $\psi^{(a)}(x,{\bf k}_T)$. It leads to a wave function
\cite{ZL}
$$\psi^{(b)}(x,{\bf k}_T)=A~(1-2x)^2~\exp\left[-\frac{{\bf k}_{T}^{2}+m^{2}}
{8 \beta^{2} x(1-x)}\right],\eqno(17)$$
and
$$\phi^{(b)}(x,1/b)==\frac{2 A \beta^2}{(2 \pi)^2}
x(1-x)(1-2x)^2\exp\left(-\frac{m^{2}}{8\beta^{2}x(1-x)}\right)
\exp\left(-2\beta^2 x(1-x)b^2\right).\eqno(17')$$
In the same way as in eq. (12), the parameters are adjusted
to be $m$=0.342 GeV, $\beta$=0.455 MeV, A=136 GeV$^{-1}$,
and $<{\bf k}_{T}^{2}>$ =(0.343 GeV)$^{2}$.

\underline{{\it Numerical calculations}}:
Numerical evaluations for the pion form factor with $\phi^{(a)}$
and $\phi^{(b)}$ are plotted in Fig. 2. We can find the perturbative
predictions are still smaller than the experimental data.
It is expected to take into account the contributions from
higher orders and higher Fock states to reach the data at the
intermediate energy.

To evaluate the effects due to the errors in the $s(\xi,b,Q)$
expression, we adopt  the
formalism of Ref. \cite {Hsiang} in our numerical calculations.
That is, we choose $t$ as defined in eq. (9)
and neglect the evolution of $\phi$ with $1/b$. In addition,
the same two models  of the distribution amplitudes in Ref. \cite{Hsiang}
are used:
(a) the asymptotic wave function \cite{LB}
$$\phi^{as}(x)=\frac{3 f_{\pi}}{\sqrt{2N_{c}}}x(1-x).\eqno(18)$$
(b) the Chernyak-Zhitnitsky wave function \cite{CZ}
$$\phi^{CZ}(x)=\frac{15 f_{\pi}}{\sqrt{2N_{c}}}x(1-x)
(1-2x)^{2},\eqno(19)$$
where $N_{c}=3$ is the number of colors and $f_\pi=0.133$ GeV
the pion decay constant.
We find that
the corrected expression (eq. (5))
increases the evaluation of the pion form factor
by  a factor of about 0.8\% for the $\phi^{as}$ and
about 1.0\% for the $\phi^{CZ}$ at
$Q=20\Lambda_{\rm QCD}$. And the effect increases with $Q$ decreasing
(reaching about 2.0\% for $\phi^{as}$ and 3.0\% for $\phi^{CZ}$ at
$Q=10\Lambda_{\rm QCD}$).
The effects are sizable individually for  the fourth and  sixth  term
in the  $s(\xi,b,Q)$ expression,
but fortunately  they cancel each other in the final expression.
As a result, the whole  effects on the pion form factor
are small.

\underline{{\it Summary}}:
In this paper, we re-examine the perturbative pion form factor with the
Sudakov suppression. It is found that in the previous formalism
there are regions where $\alpha _s$ is larger than unity and  the
perturbative predictions are still unreliable
although these regions are suppressed. Thus a cut-off
on $\alpha_s$ has to be made to guarantee the applicability
of the pQCD.
Observing that in the above regions the multi-gluon exchange
is important, we suggest that
the running coupling constant  should be
frozen at $\alpha_s(t=\sqrt{<{\bf k}_{T}^{2}>})$ when $b$ is
large and $x_1x_2Q^2$ is small by taking into account the average
transverse momentum
$\sqrt{<{\bf k}_{T}^{2}>}$. In this way, the perturbative contributions
to the pion form factor  can be calculated from the present energy
with a reasonable $\alpha_{s}$.
The essential point of our scheme is that
the running coupling constant ``frozen" is determined
by the average transverse momentum
$\sqrt{<{\bf k}_{T}^{2}>}$
which always associates  with the hadronic wave function.
In addition,
we correct the previous calculations about the Sudakov
suppression factor which plays an important role in
the perturbative predictions for the pion
form factor.

\acknowledgements{We would like to thank Dr. Hsiang-Nan Li
for his useful discussions.}

\newpage
\parindent=0pt
\section*{Figure Captions}
\begin {description}
\item [Fig. 1.]
The evolution of $\alpha_s$ with $b$ for $x_1=0.01, x_2=0.01$, $Q=2$
GeV and $\Lambda_{QCD}=100$ MeV.
The solid line is evaluated with eq. (9). The dashed
line is evaluated with eq. (10) for $\phi^{(a)}$.
\item [Fig. 2.]
The pion form factor with $\phi^{(a)}$ (solid line) and $\phi^{(b)}$
(dashed line).
\end {description}

\end{document}